\documentclass{article}
\usepackage{times}
\usepackage[hyphens]{url}      
\usepackage{multirow} 
\usepackage{rotating} 
\usepackage{longtable} 
\usepackage{graphics} 
\usepackage{color}
\usepackage{alltt}
\usepackage{rotating}

\begin{document}

\Large{A Large-Scale Study of Online Shopping Behavior} 

\begin{centering}
\large{
Soroosh Nalchigar and  Ingmar Weber\\}

Yahoo!~Research Barcelona, Barcelona, Spain\\

\{soroosh,ingmar\}@yahoo-inc.com\\
\end{centering}

\begin{abstract}
The continuous growth of electronic commerce has stimulated great interest in studying online consumer behavior. Given the significant growth in online shopping, better understanding of customers allows better marketing strategies to be designed. While studies of online shopping attitude are widespread in the literature, studies of browsing habits differences in relation to online shopping are scarce. 

This research performs a large scale study of the relationship between Internet browsing habits of users and their online shopping behavior. Towards this end, we analyze data of 88,637  users who have bought more in total half a milion products from the retailer sites Amazon and Walmart. Our results indicate that even coarse-grained Internet browsing behavior has predictive power in terms of what users will buy online. Furthermore, we discover both surprising (e.g., ``expensive products do not come with more effort in terms of purchase'') and expected (e.g., ``the more loyal a user is to an online shop, the less effort they spend shopping'') facts.

Given the lack of large-scale studies linking online browsing and online shopping behavior, we believe that this work is of general interest to people working in related areas.
\end{abstract}


\section{Introduction}\label{sec:introduction}
The continuous growth of electronic commerce constitutes a unique opportunity for companies to replace traditional ``brick and mortar'' stores with virtual ones and to reach customers more efficiently and in a larger geographical area. Online shopping as one of the types of electronic commerce has proliferated since the middle of the 1990s, aided by the parallel development of Web technologies \cite{ahn2004}. Given the business relevance of online shopping, a better understanding of customers allows better marketing strategies to be designed \cite{yang2006} and helps online retailers to beat out the increasing competition both on- and offline \cite{childers2001}. As a consequence, a growing number of studies analyze how customers use the Internet for shopping\cite{sup2007}, identifying a growing need for discovering new knowledge, models and theories on Internet customer behavior \cite{close2010}.

While studies on users' attitude concerning online shopping are widespread, studies linking online browsing habits to online shopping behavior are scarce, if existent at all. Previous works have studied various aspects of online shopping, however without paying attention to effects of Internet browsing habits on \textit{what} and \textit{how} users shop online. A comprehensive review of online shopping literature done by Chang et al.~(2005) shows that there have been no studies studying the interplay of online browsing and shopping and addressing such questions as: can general, coarse-grained browsing behavior such as the time spend on Facebook be used to predict the type of product a user will buy \cite{chang2005}? This research fills this gap by analyzing browsing data of half a million users who have bought products online from either Amazon or Walmart.

For these users we analyze (i) their pre-shopping behavior, e.g., looking at the number of related web searches or visits to product comparison sites, as well as (ii) their general, coarse-grained online browsing behavior, e.g., the fraction of page views on social networking sites or on online news portals.
Our high-level goal is three-fold.
First, we paint a detailed picture of how people shop online. Do they search before? Do they already know the store they want to go to?
Second, we test known hypotheses about offline shopping using our data. Do users spend more effort before buying expensive items?  Do they spend less effort buying items they are familiar with?
Finally, we explore the possibility to use such data for targeted advertising. Can we predict which product a user will buy based on his browsing behavior? Do users with similar browsing habits buy similar products?


The rest of this paper is organized as follows: Section 2 reviews the literature and summarizes related works within two subsections. The first part reviews related researches about online consumer behavior. The second part summarizes previous studies on online user behavior. Section 3 describes the main data source, pre-processing step, and description of data. Section 4 explains the analysis and experiments performed on prepared data and presents the results. Finally, this thesis ends with some concluding remarks in Section 5.


\section{Related Works}\label{sec:related}

In this section, we review previous related works in two areas. First, we review work that studied online consumer behavior and factors affecting it. Second, we summarize works analyzing online user behavior using ``big data'', regardless of whether related to shopping or not.

\subsection{Online Consumer Behavior}
One of the early related works to this research is done by Bellman et al. (1999). They studied the predictors of online buying behavior of 10,180 people who completed their survey that included 62 questions about online behavior and attitudes about Internet. They reported a wired lifestyle for buyers whose main characteristics are searching for product information on the Internet, receiving a large number of email messages every day, having Internet access in their offices \cite{Bellman}.

 Hasan (2010) explored gender differences in online shopping attitude.
 Data were collected from 80 students enrolled in an electronic commerce course. Results indicate a significant gender differences in cognitive, affective and behavioral components with women valuing the utility of online shopping less than their male counterparts \cite{hasan2010}.
 Close et al.~(2010) investigated the motivations of consumers' electronic shopping cart use. To gather data, these researchers recruited survey participants via an online national consumer panel. Their sample included 289 adults who have made an online purchase within the past six months. The results show that apart from immediate purchase intentions, consumers place items in their carts because of: securing online price promotions, obtaining more information on certain products, organizing shopping items, and also entertainment. They reported that only nine percent of the sample never intend to make a purchase during the same online session in which they place items in the cart, and most of them intend to purchase in the same session.\cite{close2010}. 
  Kim et al.~(2007) gathered data of 206 undergraduate students to examine the effects of image interactivity technology (IIT)  on user engagement in an  online retail environment. They showed that respondents exposed to a higher level of image interactivity, in the form of a 3D virtual model, expressed higher levels of shopping enjoyment and involvement compared to respondents exposed to a lower level of image interactivity (e.g., clicking to enlarge images), commonly used by online retailers \cite{kim}. 
Senecal et al.~(2005) performed a clickstream analysis on data of 293 participants to see how different online decision-making processes used by consumers, influence the complexity of their online shopping behavior. They reported that subjects who did not consult a product recommendation had a significantly less complex shopping behavior (e.g., fewer web pages viewed) than subjects who consulted the product recommendation\cite{senecal}.
%
Aljukhadar and Senecal (2011) performed a segmentation analysis of online shoppers based on the various uses of the internet by analyzing data of 407 participants that belonged to a consumer plan of a Canadian market research company. They found that online buyers form three segments,  the basic communicators (consumers that use the internet mainly to communicate via e-mail), the lurking shoppers (consumers that employ the internet to navigate and to heavily shop), and the social thrivers (consumers that exploit more the internet interactive features to socially interact by means of chatting, blogging, video streaming, and downloading). They concluded that online consumers differ according to their pattern of internet use\cite{aljukhadar}. 
Yang and Lai (2006) compared effects of three product bundling strategies\footnote{``Product bundling'' is a marketing strategy, examples of which include sporting organizations offering season tickets, and retail stores offering discounts when buying more than one product.} on different online shopping behaviors through a field experiment. They collected six months of log data of the behavior of 1,500 users from a publisher specializing in information technology and electronic commerce books. They indicated that significantly better decisions are made on the bundling of products when browsing and shopping-cart data are integrated than when only order data or browsing data are used \cite{yang2006}.
Hostler et al.~(2011) studied the impact of recommender systems on on-line consumer \emph{unplanned} purchase behavior. Data of this research was collected from 251 undergraduate business students. They showed that recommender systems increase product search effectiveness, user satisfaction, and unplanned purchases. 
Lee et al.~(2008) examined the effects of negative online consumer reviews on consumer product attitude. Data of their study was collected from 248 college students in Korea. They showed that negative word-of-mouth elicits a conformity effect. They found that an increase in the proportion of negative online consumer reviews causes high-involvement consumers to comply to this negative perspective. Moreover, low-involvement consumers tend to comply to the perspective of reviewers regardless of the quality of the negative online consumer reviews \cite{lee2008}.
Moon et al.~(2008) examined the influence of culture, product type, and price on consumer purchase intention for online shopping of personalized products. Data for two products,  computer desks and sunglasses, research were collected from 116 university students. The results indicate that consumers from individualistic countries were more likely to purchase customized products than those of collectivistic countries. In addition, online users are more likely to buy personalized search products than experience products. A search good is a product or service with features and characteristics easily evaluated before purchase.  On the other hand, experience goods are products or services where characteristics, such as quality are difficult to observe in advance, but these characteristics can be ascertained upon consumption\cite{luis}. We will also discuss this concept in Section \ref{sec:effcat}. Finally, they found that price did not significantly affect consumer purchase intentions \cite{moon2008}.
Verhagen and Dolen (2011) studied how beliefs about functional convenience (e.g., online store ease of use, and merchandise attractiveness) and about representational delight (enjoyment and communication style) are related to consumer impulse buying behavior. They analyzed survey data from 532 customers of a Dutch online store and showed significant effects of merchandise attractiveness, enjoyment, and online store communication style, mediated by consumers' emotions.
%
Lee et al.~(2011) studied the moderating role of social influence on online shopping and examined the impact of positive messages in discussion forums. Data of this study were collected from 104 university students in Hong Kong. They found that positive social influence reinforce the relationship between beliefs about and attitude toward online shopping, as well as the relationship between attitude and intention to shop \cite{lee2011}.
P\'{e}rez-Hern\'{a}ndez and S\'{a}nchez-Mangas (2011) analyzed the individual decision of online shopping, in terms of socioeconomic characteristics, Internet related variables and location factors. They argued that one of the relevant variables, the existence of a home Internet connection can be endogenous and have a loop of causality between variables of a model. Their dataset was from a survey conducted by the Spanish Statistical Office. Their results indicate that, compared to other variables, the effect of Internet at home on online shopping is quite small. In addition, neglecting endogeneity of Internet at home, will result in an overestimate of that variable's effect on the probability of buying online\cite{hernandez2011}.
%

Previous studies have contributed significantly and studied various aspects of online shopping, but they suffer from the following limitations:
\begin{itemize}
\item The size of datasets is typically in the hundreds, very small by web standards.
\item The data is mainly collected via questionnaires which has disadvantages such as low response rates or false replies.
\item The participants are mainly university students, which limits the generalizability of results. 
\end{itemize}

\subsection{Online User Behavior}
Over the last years, online user behavior has attracted a lot of attention, both by researchers and practitioners.
Yan et al.~(2009) examined the effects of behavioral targeting on online advertising. Their data was a log of search click behavior on a commercial search engine of 6,426,633 unique users and 33,5170 unique ads within the seven days. They reported that similar users regarding behavior on the web are likely to click on the same ads. Moreover, segmenting users for behavioral targeted advertising can significantly increase click-through rate of an ad. Finally, they found that short term user behaviors is more effective than long term user behaviors to represent users for BT \cite{yan2009}.
Though the features we use are more coarse-grained (general browsing behavior in about 25 dimensions) and though we study instances of online shopping, rather than ad clicks, we found similar trends in terms the interaction between online behavior and commercially relevant user actions.
%
%
 Kumar and Tomkins (2010) performed a large-scale study of online user behavior based on search and toolbar logs and proposed a comprehensive taxonomy of pageviews consisting of content (news, portals, games, verticals, multimedia), communication (email, social networking, forums, blogs, chat), and search (Web search, item search, multimedia search). They also studied user page to page navigation mechanisms and also the extent to which pages of certain types are revisited by the same user over time \cite{kumar2010}.
Gyarmati and Trinh (2010) performed a large-scale measurement of time spent in online social networks. They monitored 80,000 users for six weeks and found that users' total online time spent can be modeled with Weibull distributions. Also, the length of individual social networking sessions follows a power law distribution. Finally, soon after subscribing, a fraction of users tend to lose interest surprisingly fast \cite{gyarmati2010}.
%
%
 Weber and Jaimes (2010) analyzed online search behavior of 2.3 million Yahoo!~users in terms of who they are (demographics), what they search for (query topics), and how they search (session analysis). They found differences along one dimension usually induced differences in the other two \cite{ingmar2011}.
 Guo et al.~(2009), based on a Bayesian framework, proposed a click chain model and performed an experimental study on a data set containing 8.8 million query sessions. They showed that that their model outperforms previous models in a number of metrics including log-likelihood, click perplexity, and prediction of the first and the last clicked position \cite{guo2009}.
Maia et al.~(2008) clustered YouTube users to find groups that share similar behavioral patterns and reported that, as opposed to individual user attributes, user social interactions attributes are good discriminators. Their data  were collected by a web crawler from the YouTube subscription network, based on snowball sampling, and included 1,467,003 users \cite{maia2008}.
  
Although the studies above have examined various types of online user behavior (e.g., browsing behavior \cite{kumar2010}, search behavior \cite{ingmar2011}, social networking \cite{maia2008}\cite{gyarmati2010}), none of them have studied shopping behavior. The main contribution of our work is to fill this gap.\\
\\


\section{Data Set}\label{sec:dataset}
\subsection{Data Preparation}\label{sec:prepare}
We used data obtained through the Yahoo!~Toolbar as the main data source. Yahoo!~Toolbar is a browser toolbar that allows access to several functions, including Yahoo!~Search and Yahoo!~Mail. Users can optionally opt-in to give permission for Yahoo!~to log their pageviews. Basic information logged includes the timestamp, the viewed URL and, where present, its (click) referrer. URLs over https have all the dynamic parameter (such as ?q=) stripped for privacy reasons. Data obtained in this manner has been used before to study online browsing behavior \cite{kumar2010}. For our study, we used a large user-based sample of data spanning a 13 months period from February 2011 to March 2012. Note that the data did \emph{not} contain actual clear-text user IDs (such as Y!~email address) and each toolbar was simply identified by a large random number. The raw data was then processed to extract three data tables: one for users (with general browsing information), one for products (with information such as the product's approximate price), and one for shopping instances (holding information for [user,product] pairs). Using these tables, we can answer \textit{who} has bought \textit{what} and \textit{how} respectively. Data processing was doen on Hadoop using Pig as well as scripting languages. Table \ref{toy} presents an explanatory, simplified example of Yahoo! Toolbar data and included an occurrence of an online shopping. 

\begin{sidewaystable}[ht]
\centering
\small
\begin{tabular}{ccll}
\hline
\textbf{UID}&\textbf{Timestamp}&\textbf{URL}&\textbf{Description}\\\hline
fZt&1&amazon.com&User enters the shopping site\\
fZt&2&amazon.com/s/ref=nburl=search-alias\&fieldkeywords=mpe3+player&Searches for a mp3 player\\
fZt&3&amazon.com/Sony-NW-Series-Player/dp/B003WT208Q/ref=sr&Views a product page, URL has product ID and name\\
fZt&4&facebook.com/profile.php?id=1013259628&Social networking\\
fZt&5&google.com/\#hl=en\&q=cheap\%20mp3\%20players\&oq=cheap\%20mp3&Websearch about mp3 players\\
fZt&6&amazon.com/Sony-NW-Series-Player/dp/B003WT208Q/ref=sr&Clicks a search result and comes to another product page\\
fZt&7&reviews.cnet.com/17.html?query=sony\&tag=srch\&searchtype=products&Reads some reviews by other customers\\
fZt&8&amazon.com/Sony Walkman S-544/dp/B002IPHA3U/ref=lh\_ni\_t&Enters to another product page from review page\\
fZt&9&amazon.com/gp/cart/view-upsell.html?ie=UTF8\&storeID=mp3&Finally, he adds Sony Walkman to his shopping cart\\
\hline
\end{tabular}
\caption{Explanatory example of Yahoo! Toolbar data and an online shopping occurrence}\label{toy}
\end{sidewaystable}

The first step in data preparation was to filter users who have done at least some shoppings but, at the same time, who do not appear to be robots or ``mega users'' such as internet cafes. Correspondingly, we only kept ``proper'' users who, during the whole 13 months time interval, had more than 1,000 and less than 1,000,000 page views, among which there were more than 10 URLs on a large shopping sites (Amazon, Ebay, and Walmart)\footnote{Initially, we planned to include Ebay but later dropped it as a ``shopping'' event was hard to detect and due to very different characteristics for that site.}.
We further removed users whose fraction of page views on shopping sites was more than 50\%, and users located outside the  United States of America (USA).
We focus on buyers from USA  to remove effects related to differences in markets and countries, rather than by within-same-culture browsing differences. Additionally, the main language of these users is English and, consequently, most of their search queries are in English.

 At the end of this step, we are left with 485,081 users and their browsing history according to the Yahoo!~Toolbar data. The page views of each user were then divided into sessions using 30-minute time-out intervals. Thirty minutes is commonly used as threshold for breaking sessions \cite{kumar2010} \cite{zhu2009} \cite{bonchi2010}. To avoid artificial sessions that practically never time out, we also limited the maximum length to 2880 page views.\footnote{Assuming a user spends 30 seconds on each URL, visiting 2880 URLs will take 24 hours.} For a further discussion on browsing sessions, interested readers are referred to \cite{spink2006}.
 
In the next step, several processes were executed on the URL strings to extract useful information. We examined whether a given URL is a product page in Amazon or Walmart. If yes, we extracted the name of the viewed product and its ID.\footnote{Within each shop, each product has a unique identifier code.} Also, we examined if this URL corresponds to a product search on Amazon, or Walmart. The next step was to see whether the URL belongs to price comparison or product review webpages. To answer this question, we compiled a list of popular price comparison and product review websites. Next, we tried to identify the topic of a given URL (e.g., sports, games, or art). We used the existing categorization of websites from the open directory project\footnote{\url{http://www.dmoz.org}}. DMOZ' top level categories are as follows: arts (Ar), business (B), computers (C), games (G), health (H), home (Hm), news (N), recreation (R), reference (Rf), science (Scn), shopping (Sh), society (So), and sports (Sp). At the time of downloading the data, the DMOZ directory included 1,240,859 URLs. For each given URL in our browsing dataset, we iteratively truncated it from the end by removing one part at a time. We then looked for it in the DMOZ directory, and return the category(ies) to which it belongs (if any). This process repeats unil (i) a match is found or (ii) all URL suffixes/subdirectories have been stripped.

 The next step was to examine a URL to see if it belongs to a search query on either a major search engine, social networks, or on multimedia sites (e.g., YouTube). If yes, the search query was extracted from the URL (SE). Also, for each URL, we check if it belongs to social networks (S) (e.g., Facebook, Twitter, LinkedIn), multimedia (M) (e.g., YouTube, Flickr), E-mailing (Ml), and blogs (Bl) webpages\footnote{For each of these categories, we compiled and used a list of the most popular ones.}. Finally, we checked if the URL belongs to adult-content (A) using a large dictionary of the most popular such sites.

Note that we tried to collect a combination of information related to both (i) pre-shopping behavior (e.g., price comparison sites)  and to (ii) coarse-grained browsing behavior (e.g., fraction of page views on DMOZ' ``Health'' category). In the next steps, the annotated URLs are processed further to create our data tables.

\subsubsection{Users Data (Who?)}
The filtering step resulted in a set of 485,081 proper users and for each user we calculate a set of attributes that we believe are indicative of his/her Internet browsing behavior and interests. All the analysis was anonymous and performed in aggregate.  In this table, the first field is a large random number used as identifier of the toolbar instances (= users). The second field is a binary field showing that if this user is likely to be a parent. In order to label a user as potential parent, we analyzed all the URLs that they visited during 13 months, looking for some specific URL tokens, such as ``children'' or ``parenting'', which make it probable that the user is a parent. For each user, we count the number of \textit{distinct} sites (not URL) that included any of those tokens and were visited on \textit{distinct} dates. We label users as parents if they have visited 10 different sites with those tokens on 10 different days. Besides, for each user we calculate fraction of views on DMOZ categories, social networks, multimedia, web search, E-mailing, and blogs. Finally, the last attribute is total number of URLs visited by the user, which is indicative of amount of online activities of the user.

\subsubsection{Shoppings Data (How?)}\label{sec:shoppings}
This step is divided into two parts. In the first part we identify shopping intentions and in the second one, we characterize pre-shopping behavior of buyers. For each of the shopping sites, we investigated the URL sequences that are generated when a product is added to shopping cart. For us a ``shopping intention'' is defined as an instance where a user first a product view page (see Section \ref{sec:prepare}) and then shows a clear intention to pay the product.
A user shows a clear intention for payment when, from a product view page, they are either redirected (1) to a shopping cart page or (2) to a secure HTTP protocol web page (HTTPS).
To implement these definitions we use the referral URL $\rightarrow$ current URL structure to see if user is redirected from a product view page to a shopping cart page or to a secure page. This scheme ensures that, for users who are multitasking (e.g., those having multiple tabs or browser windows open), we can correctly identify the shopping intentions.
We label each expression of intent identified as a ``shopping instance'' even though we cannot be 100\% sure that, e.g., the user went through with the payment or that his credit card was accepted and so on.
 For each shopping instance, we keep the unique identifier of the user, the product ID (extracted from the product view URL), and also the timestamp and session ID. To characterize pre-shopping behavior we report a number of attributes. We count views on product pages (PView), queries issued on a shopping site (PSearch), views on price comparison (PComp) and product review (PRev) websites\footnote{These sites provide users with free services such as price comparisons, links to shopping sites, merchant ratings and customer reviews.} \emph{before} a shopping happens (within the same browsing session). Also, for each shopping instance we checked the previous search behavior of the user and count number of related queries in search engines, social networks, and multimedia sites (RelSE). Related queries were identified based on token similarity between search queries and product name, after stemming and stopwords removal. Positive similarities were considered as related search.
Moreover, we analyzed the pre-shopping search queries to find the number of times a user searched for cheap items before buying the product. To count this, we compiled a set of tokens, e.g., ``discounted'', ``cheap'', ``second hand''  that if they appear in a query, indicate that the user is looking for cheap items. Using a similar approach, we counted searches for luxury, expensive products. Finally, we recorded how users enter shopping sites before buying an item. In particular, for each shopping instance, we checked if the user has entered the shop directly. If not, we extracted the DMOZ category of the webpage from which user has transferred to shopping site, i.e., the webpage which include a link to the shoppign site. Similarly, we recorded whether a user entered the shopping site through a link on a price comparison, product review, or websearch page. If a user entered the shop from a websearch, we checked if they had issued a query that included the name of the shopping site as a token. At the end of all this processing, we are left with information for a total number of 576,209 shopping instances caused by 88,637 distinct users.

\subsubsection{Product Data (What?)}\label{sec:products}
This table keeps the product ID, name, category, and price of the products that are bought by users. To construct this table, we used the product view URLs to construct a table from product IDs to the corresponding names\footnote{A product's web address (URL) always include product uniqe code, but may or may not include a product name.}.  We found that for a given product ID there might be more than one name as, e.g., the shop may update or change the name of a given product. We removed these repetitions, and further removed those products that were not actually bought according to product IDs in the shopping table. In the end, we obtained a list of 239,491 unique pairs of product ID and names that\footnote{There were some shopping instances for which the product name was neither in the URLs viewed by the buyer, nor by the other users during the whole 13 months time span of our dataset.}. To obtain additional information for these products, such as category information or price, we queried the Yahoo!~Shopping service\footnote{\url{http://shopping.yahoo.com}} by searching for the product names\footnote{Scraping price and other product information from Amazon or Walmart would have violated the respective terms and conditions.}. To increases the accuracy rate of the Yahoo!~Shopping queries, we performed a set of cleaning steps on product names, e.g., removing tokens such as ``new'', ``good'', or ``best''.  For each product, the average of the prices of the items returned by Yahoo!~Shopping is used as a product price estimate. Also, the highest ranked category returned by Yahoo!~Shopping was used as the  product category. After obtaining the data, we inspected the accuracy of results for a set of randomly selected products. Products with names corresponding to no search results on Yahoo!~Shopping were removed. Finally, we end up with a table of 185,225 distinct products belonging to 23 different categories: Appliances (Ap), Auto Parts (Au), Babies \& Kids (B\&K), Beauty \& Fragrances (B\&F), Books (Bk), Cameras (Ca), Clothing (Cl), Computers (Co), Electronics (El), Flowers \& Gifts (F\&G), Grocery \& Gourmet (G\&G), Health \& Beauty (H\&B), Home \& Garden (H\&G), Industrial Supplies (In), Jewelry \& Watches (J\&W), Movies \& DVDs (M\&D), Music (Mu), Musical Instruments (MI), Office (O), Software (Sw), Sporting Goods (Sg), Toys (T), and Video Games (VG). 

\subsection{Data Description}\label{sec:datadescription}
This section presents descriptive statistics about our data. Table \ref{tbl:descrptive1} presents the averages of pre-shopping behavior and gives us a general picture of what buyers do before buing an item. These results are drawn from the shopping table in Section \ref{sec:shoppings} which includes 576,209 shopping instances. Table \ref{tbl:descrptive1} indicates that on average, online buyers tend to search and view products within shopping sites rather to look for them in search engines, price comparison, or product review sites.

\begin{table}[ht]
\centering
\begin{tabular}{cccccc}
\hline
\textbf{Shop}&\textbf{PView}&\textbf{PSearch}&\textbf{PComp}&\textbf{PRev}&\textbf{RelSE}\\
\hline
Amazon&12.78&8.55&0.37&0.05&0.49\\ \hline
Walmart&9.72&5.59&0.29&0.03&0.06\\\hline
\end{tabular}
\caption{Averages of pre-shopping variables for Amazon ($n=388,236$) and Walmart ($n=187,973$). See Section \ref{sec:shoppings} for abbreviations. Cases without product names are incluced here.}\label{tbl:descrptive1}
\end{table}

Table \ref{tbl:descrptive2} presents the ranking of product categories based on the percentage of bought items. To get these results, we joined the shopping and product tables on product IDs to get only those instances for which we have product information (price estimates and category). The resulting table included 303,676 shopping instances which hereafter is called shopping-product table. The results give an idea of what are highly bought categories within each shops. We see that in Amazon, Movies \& DVDs and Books are the most frequently bought items, while in Walmart Home \& Garen and  Electronics are ranked highest. Moreover, we see that for Walmart, categeories such as Babies \& Kids and also Toys are among the top ten categories, which is not the case in Amazon. Besides, findings of Tables \ref{tbl:descrptive1} and \ref{tbl:descrptive2}  indicate that the shopping instances of the two shops are quite different in terms of pre-shopping behavior and target products. Hence in the rest of this paper, to avoid side effects such as Simpson's Paradox, we analyze data of these shops separately. Also, before proceeding to next step, we add a new variable to the shopping table which is named ``effort" and its value for each shopping instance is the sum of the quantile-shifted\footnote{Each value is replaced by its percentile. E.g., a median value would be replaced by .50 and the maximum by 1.0.} normalized values of the five pre-shopping varibales mentioned in Table \ref{tbl:descrptive1}. This variable is representative of pre-shopping effort spent by the user before buying a product. Moreover, we normalize to probability distributions all the browsing variables in the user table such that for each user the sum of the DMOZ variables is 1.0. Similarly, for the sum of the other categories (social network, adult, e-mail, etc.). In the next sections we dig deeper into the data by applying various statistical and data mining techniques.

\begin{table}[ht]
\centering
\begin{tabular}{ccccc}
\hline
\multirow{2}{1cm}{\centering{\textbf{Rank}}}&\multicolumn{2}{c}{\centering \textbf{Amazon}}&\multicolumn{2}{c}{\centering \textbf{Walmart}}\\\cline{2-5}
&\textbf{Category}&\textbf{\%}&\textbf{Category}&\textbf{\%}\\
\hline
1& Movies \& DVDs&0.18&Home \& Garden&0.23\\ 
2&Books&0.13&Electronics&0.10\\
3&Home \& Garden&0.12&Clothing&0.09\\
4&Music&0.08&Computers&0.08\\
5&Computers&0.07&Babies \& Kids&0.08\\
6&Electronics&0.06&Toys&0.06\\
7&Clothing&0.04&Sporting Goods&0.06\\
8&Health \& Beauty&0.03&Video Games&0.04\\
9&Video Games&0.03&Appliances&0.03\\
10&Jewelry \& Watches&0.03&Auto Parts&0.03\\
\hline
\end{tabular}
\caption{Top ten categories in terms of percentage of shoppings for Amazon ($n=206,328$) and Walmart ($n=97,348$).}\label{tbl:descrptive2}
\end{table}


\section{Experiments}\label{sec:experiments}

\subsection{Correlation Analysis}\label{sec:cor}
In this section, the main goal is to see how users' browsing habits are correlated with their shopping behaviors. To do that, we join the shopping-product table (see Section \ref{sec:datadescription}) and the users table on the user IDs to obtain data for each shopping instance about (i) the bought product, (ii) the pre-shopping behavior and (iii) browsing behavior of the users in general. The resulting table has 303,676 rows and 50 variables\footnote{Non-numeric fields were removed for the correlation analysis.}. We calculated the Pearson correlation coefficient between all pairs of attributes and built the correlation matrix. To remove statistically insignificant correlations, we disregarged entries for $p-$values less than 0.01. From the resulting correlations we selected a set of non-obvious, more interesting rules which are relevant to the purpose of this study. E.g., we only report correlations where at least one variable related to shopping behavior. Correlations such as ``the higher the use of social networks, the lower the use of search engines'' are not discussed here.

Results indicate that for both shopping sites, being a (likely) parent is positively correlated to the number of views on product pages ($r=0.06$ in Amazon, $r=0.10$ in Walmart), to the number of searches for products before buying an item ($r=0.07$ in Amazon, $r=0.06$ in Walamrt), to number of views on price comparison sites ($r=0.02$ in both shps), and also to the total effort before shopping ($r=0.08$ in Amazon, $r=0.09$ in Walmart). Besides, we found that for both shops, fraction of views to multimedia pages (e.g., YouTube, Flickr) is positively correlated with number of product view and product searches before shopping ($r=0.02$ for both variables in each shop).

Results from Amazon show that being interested in news pages is negatively correlated with product view, product search, and the efforts before shoppings ($r=-0.05$, $r=-0.07$, and $r=-0.04$ respectively). Moreover, it is positively related to direct entering to the shopping site, i.e., without transferring from other webpages. Similarly, the fraction of views on webpages with the art topics is negatively correlated to efforts before shopping ($r=-0.04$). On the other hand, this amount is positively related to direct access to shopping sites ($r=0.04$), rather than entering the shopping sites from web search ($r=-0.04$). Other results suggest that usage of search engines is positively related to comparing prices before shopping ($r=0.04$ for both). Also it has a negative correlation with direct accessing to sites in both shopping sites ($r=-0.08$ in Amazon and $r=-0.09$ in Walmart). Other findings show that usage of multimedia sites is positively correlated to direct entering shopping sites before buying an item ($r=0.07$) and negatively correlated to entering from a web search ($r=-0.08$). Web e-mail service usage is negatively correlated to effort spent and number of product searches before shopping ($r=-0.04$, $r=-0.05$).

Results from Walmart indicate that usage of social networking sites is positively correlated with direct entering to the shopping sites ($r=0.05$) and negatively related to entering from a web serach($r=-0.06$). Moreover, sports page viewsare negatively correlated to number of product views, product searches, and also effort before shopping ($r=-0.03$ for all). Amount of using e-mail services is negatively related to shopping efforts, number of product views and alos entering shops from a websearch ($r=-0.03$, $r=-0.03$, and $r=-0.05$ respectively). Finally, regarding usage of news webpages, similar correlations to Amazon were observed in this shop.

Beside all these correlations, we found some surprising results about the interplay between effort and product price. For Amazon, we found that there is no significant correlation between them. For Walmart, a significant \emph{negative} correlation between was present ($r=-0.03$), showing the more expensive a product, the less effort is spent by buyers immediately before shopping.

To summarize the main findings of correlation: 
\begin{itemize}
\item Parents tend to finalize their shopping decision after viewing various pages, searching for products and checking prices, indicating a potential price sensitivity.
\item Regular news reader and also arts interested users have a tendency to do direct, ``impulsive'' shopping, rather to search, check prices, and read reviews before shopping.
\item Multimedia webpages users tend to have direct entering to the shopping site, rather to use a link from a web search. Also, they have a tendency to do the product view and search within the shopping sites.
\item Those buyers who have relatively high usage of search engines spend some efforts before shopping, and most probably are not impulsive buyers.
\item Usage of web-based e-mail services is related to less effort spent before shopping. 
\item A surprising relationship between the product price and the effort exists (see Section \ref{sec:PriceEfforts} for more analysis on this). 
\end{itemize}

\subsection{Product Prediction}
In this section, we use coarse-grained browsing behavior of users to predict what they will buy online. We train and test a set of classifiers for predicting what product category a user will buy based on their browsing variables. In order to remove the effect that a product can have on the browsing behavior of a user (e.g., buying a specific electronic product can change the browsing behavior in upcoming months), we built a new dataset by joining browsing behavior of the user during the first nine months and their shopping instances during the last four.

For each of the shopping sites, we prepared ten datasets that each included two different product categories. To select these categories, we tried to have five similar category pairs (e.g., books vs.~movies and DVDs) and five different categories (e.g., toys vs.~auto parts). Within each dataset, a given user had at most one shopping instance, and was never part of the same train and test set. The distribution of classes in all cases was 50\%. We used a balanced setting rather than an unbalanced one as (i) we were interested in a feasiblity study and (ii) the actual category bias would depend on the application domain. Having these datastes, we trained various classifiers and tested them using $k$-fold cross validation with $k=10$. Table \ref{tbl:classificationdetails} reports the accuracies of predictions by the Na\"{i}ve Bayes classfier in different settings\footnote{Among various classification algorithms, we found that Na\"{i}ve Bayes outperforms the rest (e.g., Support Vector Machine (SVM), $k-$nearest neighbors, and C4.5) regarding accuracy and Area Under Curve (AUC).}. 

\begin{sidewaystable}[ht]
\centering
\begin{tabular}{ccccccccc}
\hline
\textbf{Shop}&\textbf{Bk vs.~H\&G}&\textbf{B vs.~M\&D}&\textbf{Ca vs.~Cl}&\textbf{J\&W vs.~H\&B}&\textbf{S vs.~T}&\textbf{T vs.~Au}&\textbf{T vs.~B\&K}&\textbf{V vs.~T}\\
\hline
\multirow{2}{2cm}{\centering{Amazon}}&0.55&0.55&0.55&0.55&0.50&0.53&0.51&0.59\\ 
&($n=576$)&($n=2900$)&($n=566$)&($n=840$)&($n=1034$)&($n=856$)&($n=300$)&($n=760$)\\
\hline
\multirow{2}{2cm}{\centering{Walmart}}&0.500&0.52&0.52&0.55&0.53&0.52&0.55&0.54\\
&($n=372$)&($n=151$)&($n=816$)&($n=454$)&($n=1468$)&($n=766$)&($n=1155$)&($n=1283$)\\
\hline
\end{tabular}
\caption{Details of product category prediction based on users' coarse-grained browsing features. See Section \ref{sec:products} for abbreviations.}\label{tbl:classificationdetails}
\end{sidewaystable}

Results indicate that for Amazon, we have 59\% accuracy in predicting what category a user will buy among the similar categories video games and toys. This value is 55\% for books vs.~home \& garden, books vs.~music \& DVD, and also cameras vs.~clothing. For Walmart, we observed 55\% accuracy in prediction of jewelery \& watches vs.~health \& beauty, and the same accuracy for predicting toys vs.~baby \& kids. It should be remarked that we are using high level chatacteristics of users to predict a low level feature such as the product category that they will buy online. Given that we use only coarse-grained browsing features, the noticeable and statistically significant improvement over the random baselines indicates the plausibility of the idea that online browsing behavior could be used for product prediction and to improve product recommendations. 


\subsection{Clustering of Buyers}
\begin{figure}[ht]
\begin{minipage}[b]{0.5\linewidth}
\centering
\includegraphics[scale=0.48]{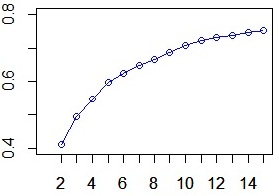}
\caption{Elbow method for Amazon.}\label{fig:elbow:amazon}
\end{minipage}
\begin{minipage}[b]{0.5\linewidth}
\centering
\includegraphics[scale=0.5]{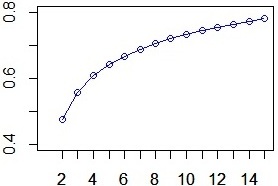}
\caption{Elbow method for Walmart.}\label{fig:elbow:walmart}
\end{minipage}
\end{figure}

The goal of this section is to discover clusters of online buyers and characterize them based on their Internet browsing habits and the products  bought. We use the $k-$means clustering algorithm but, conceptually, any clustering algorithm could have been chosen. The initial dataset that we used in this section was the same as in Section~\ref{sec:cor} which includes 303,676 shopping instances caused by 88,637 distinct users. We split this dataset based on shopping sites and, for each site, constructed a set of distinct users. The total number of distinct users for Aamzon and Walamrt are 63,641 and 34,235 respectively with some users being in both datasets. Due to normalization performed we chose not to include the (binary) parent variable in the clustering setup. To choose an appropriate number of clusters of buyers for each shop, we implemented the so-called ``Elbow method''. Figures \ref{fig:elbow:amazon} and \ref{fig:elbow:walmart} show the results in which the $x$ axis is number of clusters and the $y$ axis is percentage of between cluster distance out of total distance. Each data point corresponds to an average over 50 runs of $k-$means with random initializations.

Based on these figures we chosed $k=5$ for Amazon and $k=4$ for Walmart. Then, for each shop, we performed the $k-$means clustering method 100 times from which we kept the one with the highest percentage of between cluster distance out of total distance. Table \ref{ClusAvgMed} presents the size of clusters and shows the average effort and the medians price spent by users within the clusters. Additionally, Table~\ref{tbl:clusteringdetails} presents results about browing behavior and the top product categories for each of the clusters. 

\begin{table}[ht]
\centering
\begin{tabular}{ccccc}
\hline
\textbf{Shop}&\textbf{Cluster}&\textbf{Size}&\textbf{Avg. Efforts}&\textbf{Med. Price}\\
\hline
\multirow{5}{1.5cm}{\centering{Amazon}}&1&6,432&1.408&35.99\\
&2&20,525&1.405&30.98\\
&3&13,522&1.449&31.39\\
&4&15,378&1.401&31.00\\
&5&7,784&1.295&30\\
 \hline
\multirow{4}{1.5cm}{\centering{Walmart}}&1&14,566&1.071&69.00\\
&2&9,212&1.106&66.31\\
&3&4,911&1.054&71.91\\ 
&4&5,546&1.089&68.80\\
\hline
\end{tabular}
\caption{Size, average effort spent and median prices for different clusters.}\label{ClusAvgMed}
\end{table}

\begin{sidewaystable}[ht]
\centering
\small
\renewcommand{\tabcolsep}{0.2cm}
\renewcommand{\arraystretch}{1}
\begin{tabular}{ccll}
\hline
\textbf{Shop}&\textbf{Cluster}&\textbf{Browsign habits of centroids (\%)}&\textbf{{Product categories}}\\
\hline
\multirow{5}{1.5cm}{\centering{Amazon}}&1&M $0.45^+$ , Bl $0.01^+$ , Ar $0.04^+$ , G $0.01^+$ , Scn $0.006^+$.&H\&G $\uparrow$, B $\downarrow$, C $\uparrow$, M $\downarrow$, VG $\uparrow$,Cl $\downarrow$ Sg $\uparrow$.\\
&2&S $0.79^+$ , G $0.008^-$, Sh $0.08^-$, Sp $0.007^-$.&E l$\uparrow$, C o$\downarrow$, VG $\uparrow$, Cl $\downarrow$, H\&B $\downarrow$, Sg $\uparrow$.\\
&3&Sh $0.44^+$ , N $0.06^+$,  B $0.09^+$ , H $0.007^+$, Hm $0.03^+$,&H\&G $\uparrow$, Bk $\downarrow$, H\&B $\uparrow$, Cl $\downarrow$, J\&W $\uparrow$, Sg $\uparrow$.\\
& & R $0.02^+$, Rf $0.02^+$, Scn $0.008^+$, SE $0.25^+$, So $0.03^+$.&\\
&4&G $0.01^+$ , H $0.005^+$ , Hm $0.02^+$ , R $0.02^+$, Scn $0.006^+$, So $0.03^+$.&J\&W $\uparrow$, H\&B $\downarrow$, VG $\downarrow$.\\
&5&S $0.11^-$ , A $0.10^+$ , Ml $0.25^+$ ,  N $0.2^+$ , Ar $0.05^+$ , B $0.09^+$ , G $0.01^+$,&J\&W $\uparrow$, Sg $\uparrow$.\\
 &&H $0.005^+$ , Hm $0.03^+$ , R $0.02^+$ , Scn $0.007^+$, So $0.03^+$ , Sp $0.02^+$.&\\
 \hline
\multirow{4}{1.5cm}{\centering{Walmart}}&1&S $0.83^+$ , A $0.009^+$ , B $0.03^-$ , Hm $0.008^-$, Sh $0.07^-$. &B\&K $\uparrow$,  El $\downarrow$, Cl $\downarrow$, Co $\downarrow$, Au $\uparrow$, Ap $\downarrow$.\\
&2&A $0.01^+$, G $0.01^+$, H $0.004^+$, Hm $ 0.021^+$,  Scn $0.005^+$&Cl $\uparrow$, El $\downarrow$.\\
&3&S $0.17^-$, A $0.08^+$, M $0.21^+$, Ml $0.14^+$, N $0.08^+$, Ar $0.04^+$,&Co $\uparrow$, Cl $\downarrow$, Sg $\uparrow$, T $\downarrow$.\\
&&Hm $0.022^+$,  Scn $0.005^+$, Sp $0.01^+$&\\ 
&4&S $0.12^-$, A $0.01^+$, SE $0.66^+$, N $0.06^+$, B $0.1^+$, Hm $0.031^+$, &Ap $\uparrow$, VG $\downarrow$, H\&B $\uparrow$.\\
&&R $0.02^+$, Rf $0.02^+$, Scn $0.005^+$, Sh $0.26^+$, So $0.03^+$&\\
\hline
\end{tabular}
\caption{Details of clustering of users based on their browsing habits. The $^+$ and $^-$ indicate that a certain feature is under- or overexpressed compared to the corresponding cluster-independent first and third quartile of the feature. The $\uparrow$ and $\downarrow$ show the changes in the ranking of the top ten categories bought within each cluaster compared to a cluster independent category ranking. See Section for \ref{sec:prepare}  abbreviations of browsing categories and Section \ref{sec:products} for abbreviations of product categories.}\label{tbl:clusteringdetails}
\end{sidewaystable}

These results indicate that in Amazon, the first cluster mainly includes users that tend to spend a lot of time on multimedia pages, as well as reading blogs, and also have a slight tendency towards arts, games, and science. For these users, categories such as home \& garden, computers, video games, and sporting goods are more popular than for overall Amazon users. The second cluster is formed by social network users which tend to use more the internet's interactive features to communicate with others. For these buyers, the fraction of views on game and shopping tend to decrease and prduct categories such as electronics and video games place at higher ranks. Surprisingly, we found that although for these users views on sport pages tend to decrease, the sporting goods category has a higher rank. The third cluster are mostly shoppers, those users that employ internet to navigate and browse shopping sites and related pages. Besides, they have a relatively high fraction search services page views, and also views on business related pages. For these users, we observed that categories home \& garden, helath \& beauty, and jewelry \& watch have higher ranks. It should be noted that in comparison to other clusters, these users tend to spend higher amount of efforts before shopping. The forth cluster of Amazon buyers are formed by buyers whose internet browsing include more health, home, and society related webpages. The ranking of top ten product categories for these users seems to be similar to the cluster independent ranking, except for the jewelry \& watch category which is higher and for health \& beauty as well as video games which is lower. The last cluster of Amazon seems to be formed by users that have various interests and use internet for different purposes, among others for e-mailing, news reading, and adult-content views. For these users, jewelry \& watch, and sporting goods categories have higher rank. On average, users from this cluster spent the least shopping effort and tend to buy cheaper products (based on median prices within the cluster) than shoppers from other clusters (See Table \ref{ClusAvgMed}).

For Walmart, we found similar clusters to Amazon, indicating that the general browsing behavior does not depend too much on which online shop a user frequents. The first cluster is formed by social networkers. Categories such as babies \& kids, and auto parts have a higher rank here. The second cluster of Walmart shoppers include those who seem to be general users, who use internet for various purposes, e.g., home, health, and science. For these people, clothes has a higher rank and electronics lower. 
Buyers from this category, on average, tends to spend the highest shopping effort and pay the lowest price (based on median prices within the cluster). The third cluster includes mostly multimedia and e-mail users. For these users, sporting goods have higher rank and clothings and toys are lower. Buyers from these cluster, on average, have paid the highest prices (based on median prices within the cluster) and spent the least effort, similar to the first and last clusters of Amazon. The last cluster include web searchers and shoppers. For these buyers, categories appliance and health \& beauty has higher rank and video games places in a lower position.
Results presented in this section are in accordance with \cite{aljukhadar}\footnote{In comparison to that work, our study was based on large-scale data and resulted in more detailed clusters.} where the authors give a 
segmentation of online shoopers, and could be used to design/improve marketing campaigns and to better target online shoppers segments based on their Internet browsing habits. It should be mentioned that although the clusters within two shops seems very similar, we do not expect their shopping categories to be simialr, due to differences in the shops' focus (see Section \ref{sec:datadescription} for further discussion). 

\subsection{Product Categories and Effort}\label{sec:effcat}
The goal in this section is to see how the amount of efforts spent before shopping differs for various product categoris. Towards this end, we calculate the average effort for each category and compare it with the category-independent  average of efforts. Dataset that we use in this section is same as Section \ref{sec:cor}. We report the results that are statistically significant ($p-$value < 0.01) and include at least a 5\% relative change in the macro average.  We found that within both shops, the category home \& garden increases the efforts by 5\% in each shop and the category health \& beauty increases the efforts by 5\% in Amazon and 9\% in Walmart. Moreover, we found that the software category has an 8\% lower effort in each shop.

Additionally, we found that in Amazon categories such as auto parts, jewelry \& watches, clothing, and musical instruments increase the efforts by 5\%, 5\%, 7\%, and 8\% respectively. In addition, categories such as books, movies \& DVDs, and video games decrease effrots by 6\%, 7\%, and 5\%. For Walmart we found that categories such as beauty \& fragrances and office increase efforts by 5\% and 15\% and the category flowers \& gifts decrease the efforts by 32\%.

 The results of this section, in accordance with results of \cite{moon2008} and \cite{gretzel2002}, show that online consumers behaviors differs based on the type of the prodcut on which they are making decison. In particular, ``experience goods'' correspond to more effort and difficulties for online buyers in accurately making online shopping decision since at the moment of shopping only (abstract) information about the product, but not the product itself, is available. On the other hand, ``search products'' are asscoiated with less effort and inspection before shoppings.
  Experience goods are products or services where characteristics, such as quality are difficult to observe in advance, but these characteristics can be ascertained upon consumption. On the other hand, a search good
is a product or service with features and characteristics easily evaluated before purchase \cite{luis}.
Our experiments indicate that for experience goods, such as appliances or cameras, online buyers tend to spend more effort before shopping, in contrast to search goods, such as software or flowers \& gifts.

\subsection{Customers Loyalty and Effort}
\begin{figure*}[ht]
\centering
\begin{minipage}[b]{0.5\linewidth}
\centering
\includegraphics[scale=0.7]{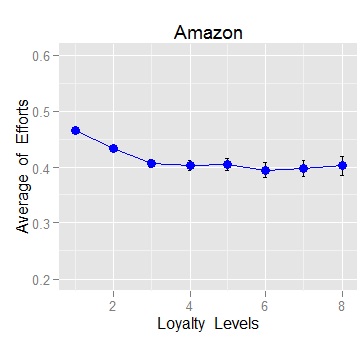}
\caption{Shopping effort drops with an increase of loyalty to Amazon.}\label{fig:adaptation:amazon}
\end{minipage}
\begin{minipage}[b]{0.5\linewidth}
\centering
\includegraphics[scale=0.7]{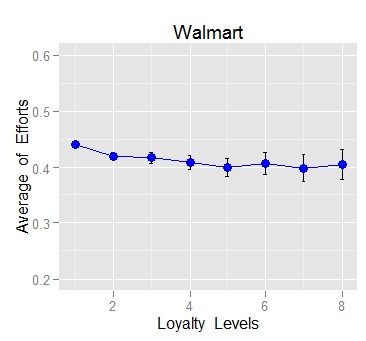}
\caption{Shopping effort drops with an increase of loyalty to Walmart.}\label{fig:adaptation:walmart}
\end{minipage}
\end{figure*}

Success of online shops depends largely on customer satisfaction and other factors that will eventually increase customers' loyalty\cite{loyalty1}. In this section, the main goal is to examine the existence of loyalty intentions toward online shopping sites and to investigate how pre-shopping effort changes for different levels of loyalty. In particular, we want to see how pre-shopping efforts changes as customers get used to the shopping sites and their loyalty increases. To measure customer loyalty, we use one of the early, and widely used definitions which is repeated purchasing \cite{loyal2}\cite{loyal3}\cite{loyal4}. The dataset we use in this section is the same as in previous sections, only that now we count and accumulate the ``loyalty level'' of a shopping instance for a user on either Amazon or Walmart. An level of 1 corresponds to the first time the user buys and item on that shop during our 13 months window. A level of 2 indicates the second time and so on. Using this data, we plot the effort for different loyalty levels and examine the trend. Figures \ref{fig:adaptation:amazon} and \ref{fig:adaptation:walmart} show the results for Amazon and Walmart respectively. In these figures, each datapoint is the average of the effort spent by buyers for the corresponding loyalty level and includes at least 400 instances. Black lines are interval estimation (with 95\% confidence) for the mean effort at the corresponding loyalty level.

Results indicate that increasing the loyalty level is negatively correlated to the amount of efforts that buyers spend before shopping. In other words, as users buy more products from a shopping site, they tend to lower the amount of effort spent on shopping. This could be due to increased trust to the onlien shoppign site, and also effets of learning how to shop online. Finally, we mention that we performed a similar analysis for each product categories separately and we found similar, downwards trends, only with larger error bars due to the smaller number of samples.

\subsection{Price and Efforts}\label{sec:PriceEfforts}
\begin{figure*}[ht]
\begin{minipage}[b]{0.5\linewidth}
\centering
\includegraphics[scale=0.7]{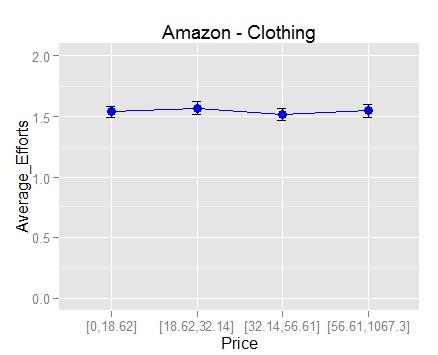}
\caption{Effort and price for clothign product in Amazon}\label{fig:PEamazon1}
\end{minipage}
\begin{minipage}[b]{0.5\linewidth}
\centering
\includegraphics[scale=0.7]{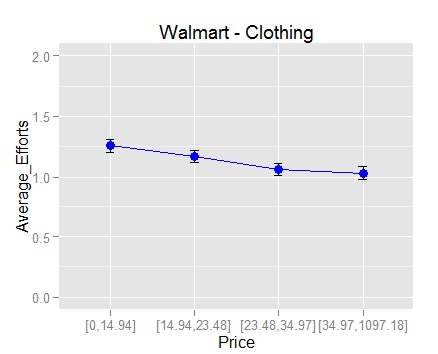}
\caption{Effort and price for clothign product in Walmart}\label{fig:PEwalmart1}
\end{minipage}
\end{figure*}

As we saw in Section \ref{sec:cor}, there are some surprising correlations among price and pre-shoppings. In this section, we dig deeper by examing the changes in pre-shopping efforts for different levels of price within each of the categories separately. We use  four quantiles (quartiles) to set breaking point for the price axis. Within each category and for each of the price buckets, we calculate the average effort spent on shopping instances and estimate a 95\%-confidence interval for the mean. Figures \ref{fig:PEamazon1} and \ref{fig:PEwalmart1} show the results for Amazon and Walmart for the clothing category. Our analysis gave similar plots for other product categories within the same shop. We see that for Amazon, the amount of effort tends to be stable as the price changes. However, for Walmart, a negative correlation is observed, which means that shoppers in this shop, tend to spent less shopping effort for more expensive products. We also calculated the correlation coefficient between price and effrots for each of the categories which resulted in significant negative correlations in Walmart (e.g., $r=-0.09$ in electronics, $r=-0.14$ in computers, and $r=-0.08$ for babies \& kids). However, there is no significant correlation for Amazon.

\section{Conclusions}\label{sec:conclusions}
Explosive growth of electronic commerce and the increasing number of online users has caused increasing interest in online consumer behavior. Understanding online shopping behavior and factors that affect it is important for researchers and practitioners alike. We use a large sample of 13 months of user browsing logs to create three data tables for (i) users, (ii) products, and (iii) shopping instances. Using various statistical and data mining techniques, we mined these tables to discover interesting patterns from them. We found various significant correlations between internet browsing features of buyers and their pre-shopping behavior. We showed that these coarse-grained internet browsing features of online shoppers could be used to predict what product category a user will buy with a noticeable improvement over a trivial baseline. We also showed that online consumers could be segmented into different clusters based on their internet browsing habits. We characterized such clusters and found that clusters are different regarding product categories and other characteristics such as price and shopping effort. Additionally, we found that the amount of effort that online consumers spend before buying an item differs for various product categories. Results suggest that experience goods are associated with more effort in making buying decision while, on the other hand, search products are associated with less effort. Moreover, we found that an increase in the level of shop loyalty of users comes with a decrease in the effort that users spend before shopping. Surprisingly, we did \emph{not} find the expected relationship that more expensive items would come with more shopping effort. This could be due to long-term shopping behavior in which a user plans a purchase and spends extensive shopping efforts for it outside of the web, or it could be an indication that people buying such products are also less price sensitive.

This is the first work we are aware of that analyzes hundreds of thousands of online shopping instances and links them to general browsing behavior of the same user. Though our work is primarily of descriptive nature, we believe that given the important of online shopping a systematic data desription such as it is presented here has intrinsic value for people working in related areas. In the presence of additional data sources such as email or instant messenger networks, our study could be extended further to incorporate social networking information. We leave such an extension to future work.



\bibliographystyle{abbrv}
\bibliography{online_shopping}  

\end{document}